\documentclass[
reprint,
 amsmath,amssymb,
 aps,
pra,
floatfix,
]{revtex4-2}
\usepackage{hyperref}
\usepackage{graphicx}
\usepackage{dcolumn}
\usepackage{bm}
\usepackage{braket}
\usepackage{amsmath}
\usepackage{physics}
\usepackage{amsfonts}
\usepackage{xspace}
\usepackage{color}
\usepackage{xcolor}
\usepackage{amssymb}
\usepackage{hyperref}

\begin{document}

\newcommand{\jjh}[1]{\textcolor{red}{#1}}
\newcommand{\sah}[1]{\textcolor{blue}{#1}}
\newcommand{\eq}[1]{Eq.~(\ref{#1})\xspace}

\preprint{APS/123-QED}

\title{Effects of Systematic Error on Quantum-Enhanced Atom Interferometry}
\begin{abstract}
We develop a framework for describing the effects of systematic state preparation error in quantum-enhanced atom interferometry on sensing performance. 
We do this in the context of both spin-squeezed and non-Gaussian states for the two-axis-twisting (TAT), one-axis-twisting (OAT), and twist-and-turn (TNT) state preparation schemes, and derive general conditions for robustness and susceptibility of quantum states to state preparation error. In the spin-squeezing regime, we find that OAT is more susceptible to state preparation error than TAT due to its parameter-dependent phase space rotation. In the non-Gaussian regime, we find that OAT is robust to state preparation errors, which can be explained by a small ratio of off-diagonal to diagonal elements in its Fisher-covariance matrix. In contrast, TNT does not exhibit this robustness. We find that the single parameter unbiased estimators that are habitually used in quantum-enhanced atom interferometry are not always optimal, and that there may be occasions where biased estimators, or two-parameter unbiased estimators, lead to lower net error.

\end{abstract}
\author{Joshua Goldsmith}
\email{Joshua.Goldsmith@anu.edu.au}
\affiliation{Department of Quantum Science and Technology, Research School of Physics, The Australian National University, Canberra, Australia}
\author{Joseph Hope}
\affiliation{Department of Quantum Science and Technology, Research School of Physics, The Australian National University, Canberra, Australia}
\author{Simon Haine}
\affiliation{Department of Quantum Science and Technology, Research School of Physics, The Australian National University, Canberra, Australia}
\maketitle
\section{Introduction}
Atom interferometers are measurement instruments relying on the interference of matter-waves \cite{atomint,ramtransatmint}, and are currently used for precision sensing of a wide variety of quantities, including magnetic fields \cite{frequselectmag, Vengalattore:2007}, accelerations \cite{Canuel:2006}, rotations \cite{Gustavson:1997},  gravitational fields \cite{Peters:2001, Hu:2013, Canuel:2018}, and gravity gradients \cite{Snadden:1998}. Unlike classical devices, which may drift over time and thus require regular calibration, the response of an atom interferometer can, in principle, be locked to fundamental constants of nature \cite{Geiger2011,Geiger:2020, Canuel:2018}, making it inherently robust against long-term drift. This key advantage of atom interferometry enables highly sensitive precision measurements, such as those of the fine-structure constant \cite{Parker:2018} and Newton's gravitational constant \cite{Rosi:2014}.

There is considerable recent interest in using quantum entanglement to increase the sensitivity of atom interferometry through reducing the atomic shot-noise \cite{Pezze_review:2018}. Holding everything else equal, reducing the atomic shot-noise would improve the per-shot sensitivity, increasing the rate at which a given sensitivity is reached when integrating down. This would ultimately improve the measurement bandwidth, allowing for more precise measurements of slowly changing signals, or allow for higher sensitivities when the experiment is limited by drift in other quantities \cite{Szigeti:2020,Szigeti:2021}. Alternatively, incorporating quantum-entanglement while holding precision fixed would allow for smaller interrogation times, leading to significantly smaller devices \cite{Szigeti:2021}, and devices less prone to external perturbations \cite{Wang:2023}. There have been several proof-of-principle demonstrations of entanglement-enhanced atom interferometry, with entanglement generated via either atom-atom \cite{Gross:2010, Riedel:2010, Lucke:2011, Hamley:2012, Berrada:2013, Strobel:2014, Muessel:2014, Muessel:2015, Kruse:2016, Zou:2018, Laudat:2018} or atom-light \cite{Appel:2009, Leroux:2010, Schleier-Smith:2010, Sewell:2012, Hosten:2016, Greve:2022} interactions. Several alternate schemes have also been theoretically proposed, but are yet to be experimentally demonstrated \cite{Hammerer:2010, Haine:2011, Haine:2013, Haine:2014, Haine:2016, Szigeti:2017, Lewis-Swan:2018, Haine:2020, Fuderer:2023, Barberena:2024}.

While there have been many studies on how quantum-enhancement affects the \emph{precision} of atom interferometers, there has yet to be an investigation on how it affects their \emph{accuracy}. In particular, the key advantage of atom interferometry is that the response only depends on fundamental constants, which are well-known and static. Once atom-atom, or atom-light interactions are introduced into the scheme, the response now depends on less easy to characterise and control parameters. For example, the one-axis twisting scheme \cite{Kitagawa1993} can be implemented via atom-atom interactions \cite{Haine:2009, Haine:2014}. Deviations in any number of parameters, such as the atomic scattering length or even the trapping configuration, will affect the initial state and ultimately the response of the device. Similarly, atom-light interactions could be affected by parameters such as laser power, or even the size of the beam focus, both of which are susceptible to drift \cite{Kuzmich2000, Kritsotakis:2021}. These considerations raise the question: Given that the stability of calibration is the key advantage of atom interferometry, do the added complexities of entanglement-generation schemes outweigh the precision gains they offer?

In this paper, we investigate the impact of systematic errors in entangled state preparation on the theoretical precision and accuracy of atom interferometers. Through investigating the one-axis-twisting (OAT) and two-axis-twisting (TAT) processes, we consider systematic state preparation errors in spin-squeezed states, and demonstrate how in some circumstances, biased estimators may meaningfully increase parameter resolution. We examine the degree of state preparation error in OAT that would degrade quantum enhancement by various factors for metrologically useful atom ensemble sizes, providing a schema for calculating the effect of entangled state preparation error for an entangled atom interferometry sequence.  We then consider errors in state preparation in the non-Gaussian regime through the OAT and twist-and-turn (TNT) state preparation schemes. In both the spin-squeezed and non-Gaussian regimes, we identify general classes of quantum states that are resistant to state preparation error. We present a means of overcoming the biasing effects of state preparation error using dual parameter maximum likelihood estimation.

\section{Quantum-enhanced atom interferometry and parameter estimation}
\begin{figure*}
    \centering
    \includegraphics[width=1.0\textwidth]{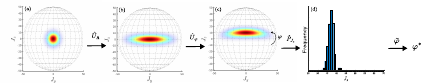}
    \caption{A quantum-enhanced atom interferometry sequence. In (a), an initial coherent-spin state, entangled in (b) according to $\hat{U}_{\mathbf{\Lambda}}$, with the parameter $\varphi$ encoded onto the state in (c). In (d) a projective measurement in the $\hat{J}_z$ basis is made, with an estimate $\varphi^*$ occurring based on the measurement results according to an estimator $\Tilde{\varphi}$.  }
    \label{fig:schema}
\end{figure*}

An atom interferometry sequence can be divided into four stages: state preparation, parameter encoding, measurement, and parameter estimation, as depicted in Figure \ref{fig:schema}. 
Quantum enhancement occurs in the state preparation stage, where correlations are induced in an initially unentangled atomic ensemble to surpass the shot-noise limit. The preparation of a metrologically useful entangled state, and the subsequent encoding of a parameter $\varphi$ of interest, can be represented as
\begin{equation}
    \label{eqn:genunitary}
    \ket{\psi_{\varphi, \mathbf{\Lambda}}}=\hat{U}_\varphi\hat{U}_{\mathbf{\Lambda}}\ket{\psi_0},
\end{equation}
where $\mathbf{\Lambda}$ represents a vector of state preparation parameters. In this paper, we only consider state preparation schemes with one degree of freedom, which we denote as $\lambda$. All subsequent state preparation parameters are either constant or a function of $\lambda$. As an example, the OAT spin-squeezing scheme relies on a nonlinear interaction between the atoms to generate entanglement. In this case, the strength of that interaction would be the $\lambda$ state preparation degree of freedom in $\mathbf{\Lambda}$, with the subsequent phase space rotation a $\lambda$-dependent state preparation parameter. After the parameter $\varphi$ is encoded onto the system, a measurement $\hat{P}$ is made in some basis, generating a set of measurement results $\mathbf{X} = \{X_1, ...,X_n\}$, which are used to estimate $\varphi$ via a function $\Tilde{\varphi}$, termed an estimator, producing an estimate $\varphi^* = \langle \Tilde{\varphi}\rangle$. For a given estimator, the uncertainty in the estimate is given by the mean-square error:
\begin{align}
\text{MSE} &= \langle\left(\Tilde{\varphi} - \varphi\right)^2\rangle \notag \\
&= \sigma_Q^2 + \sigma_B^2,
\end{align} 
where
\begin{align}
\sigma_Q^2 &= \langle \Tilde{\varphi}^2\rangle - \langle \tilde{\varphi}\rangle^2 \notag \\
&= \langle \Tilde{\varphi}^2\rangle - \left(\varphi^*\right)^2
\end{align}
is the contribution from sampling error and statical fluctuations resulting from the quantum projection noise in the measurement $\hat{P}$, and 
\begin{align}
\sigma_B &= \langle \left(\Tilde{\varphi} - \varphi\right)\rangle \notag \\
&= \left(\varphi^* - \varphi\right)
\end{align}
is the systematic error, or bias. The MSE is the total mean-squared distance between our estimate of the parameter and its true value, and is an overall metric for error accounting for both bias in a parameter estimate, and fluctuations around that parameter estimate.

For an unbiased estimator ($\sigma_B =0$),  the optimal precision of a readout distribution is given by the Cramer-Rao bound (CRB)
\begin{equation}
    \label{eqn:CRB}
     \text{MSE} = \sigma_Q^2 = \frac{Q^2}{m} \geq \dfrac{1}{m F_c},
\end{equation}
where $m$ is the number of repetitions of the experiment, $Q$ is the single-shot quantum projection noise, and $F_c$ is the classical Fisher information, given by
\begin{equation}
\label{eqn:ClassicalFisher}
    F_c=\sum_j \frac{(\partial_\varphi P_\varphi(j))^2}{P_{\varphi}(j)}
\end{equation}
for measurement values $j$ and a probability distribution $P(j)$ \cite{Fischer1922}. The estimator $\Tilde{\varphi}$ that is guaranteed to saturate the CRB for all distributions is the maximum likelihood estimator, but other classes of estimators may also fulfil this constraint in particular scenarios. The fundamental limit to single parameter quantum sensing is achieved through saturating the Quantum-Cramér-Rao bound (QCRB): $F_c = F_Q$, where $F_Q$ is the quantum Fisher information \cite{Braunstein:1994, Pezze_review:2018}.
 
The minimally noisy unbiased estimator for a given state preparation sequence is a function of the degree of entanglement in the state, and hence on the state preparation degree of freedom. The estimator parameterised by the state preparation degree of freedom $\lambda_p$ is denoted $\tilde{\varphi}(\mathbf{X}| {\lambda_{p}}$) for a set of measurement data $\mathbf{X}$.

\subsection{Effects of state preparation error}
\begin{figure}
    \centering
    \includegraphics[width=\columnwidth]{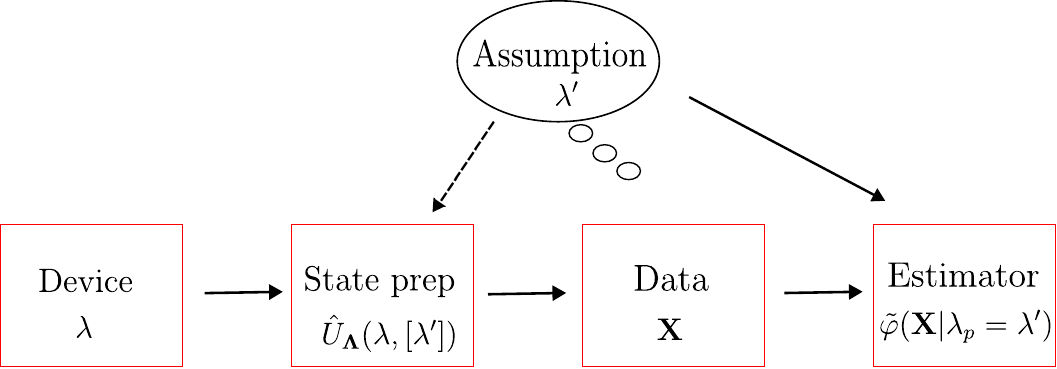}
    \caption{A depiction of how the set of state preparation parameters $\mathbf{\Lambda}$, and the actual, assumed, and estimator state preparation degrees of freedom ($\lambda,\lambda', \lambda_p$) interact to give error in state preparation. The real $\mathbf{\Lambda}$ values totally characterise the measurement data $\mathbf{X}$, and the value $\lambda_p$ is determined by $\lambda'$, which in turn is used to process the measured data to obtain a parameter value with a particular mean squared error. Assuming a particular state preparation degree of freedom may result in later parameters in the state preparation sequence being suboptimal, hence the dashed line.}
        \label{fig:lam_fig}
\end{figure}

State preparation error occurs when we \emph{believe} that we have prepared the state 
\begin{equation}
    \label{eqn:errorunitary}
    \ket{\psi_{\varphi, \mathbf{\Lambda'}}}=\hat{U}_\varphi\hat{U}_{\mathbf{\Lambda'}}\ket{\psi_0} \, ,
\end{equation}
but have actually prepared the state in Equation \ref{eqn:genunitary}, with $\mathbf{\Lambda} \neq \mathbf{\Lambda'}$. We use the notation that the dashed superscript denotes that a variable is assumed, while the absence of a dash implies the actual physical value. Given a state preparation error, the $\lambda_p$ value used in our estimator will be $\lambda_p=\lambda'$ instead of $\lambda_p=\lambda$, as this will correspond to the estimator we \textit{think} will be unbiased. As our estimator is then parameterised by the wrong value, it will output an expected value $\langle \Tilde{\varphi}(\mathbf{X} | \lambda') \rangle=\varphi^*$, where the true phase value $ \varphi \neq \varphi^*$ in most instances. In addition to the introduction of bias, the biased estimator will lead to different single shot quantum noise $Q$ relative to an unbiased estimator. Furthermore, assuming the wrong state preparation degree of freedom \textit{may} bias other state preparation parameters dependent on $\lambda$, thus changing the measurement data as well as the estimator. These distinct measurement data may cause the state preparation error to further change $Q$ relative to an unbiased quantum-enhanced atom interferometry sequence. Figure \ref{fig:lam_fig} demonstrates how each of the actual, assumed, and estimator state preparation values interact to produce a value for the error in the parameter of interest. 

In many cases, such as when using spin-squeezed states, bias induced by state preparation error will be linearly proportional to the parameter $\varphi$ for $\varphi \ll 1$, and the total MSE will be characterised by the parameters $B$ and $Q$:
\begin{equation}
    \text{MSE} = (B \varphi)^2 + \frac{Q^2}{m} \, , \label{MSE2}
\end{equation}
where $B$ is defined by $\sigma_B = B \varphi$. As the bias decreases when operating close to the bias-free operating point of $\varphi = 0$, it is desirable to operate as close to this point as possible. Adaptive schemes will be ultimately limited by the resolution of the measurement. As such, we can obtain a self-consistent lower bound on the MSE by setting $\varphi = Q/\sqrt{m}$ in Equation \ref{MSE2}, which yields $\text{MSE} \geq E^2/mN$, where
\begin{align}
E &= \sqrt{N Q^2(1+B^2)} \, \label{eqn:errormetric}
\end{align}
is the total error of the measurement, normalised by the precision attainable in a single bias-free, uncorrelated atomic ensemble. Alongside $B$ and $Q$, $E > 1$ is a useful indicative quantity as it implies that the presence of state preparation error has worsened the precision of atom interferometry relative to using an unentangled state - one would have been better off dispensing with entangled state preparation entirely. 

\section{Systematic state preparation error in spin-squeezed states}
We now turn our attention to the specific example of spin-squeezed states. In this paper, we confine the states explored to two-mode bosonic states. We describe our quantum state using second quantised notation, and define a set of pseudo-spin operators 
\begin{equation}
    \label{eqn:pseudospinoperators}
    \hat{J}_k = \frac{1}{2}(\hat{a}^{\dagger}_1\hat{a}^{\dagger}_2) \sigma_k (\hat{a}_1 \hat{a}_2)^{T}
\end{equation}
acting on our state, where $\sigma_k$ corresponds to the ${k}^{\text{th}}$ Pauli matrix, and $\hat{a}_1$ and $\hat{a}_2$ represent the first and second annihilation operators for a Fock space respectively \cite{B.Yurke1986}. The $\hat{J}_z$ operator corresponds to half the number difference of the two modes the atoms can populate, and is usually our physically accessible measurement. Certain classes of two-mode bosonic states can yield subshot noise sensitivity based on moments of a readout distribution. An important example of this is spin-squeezed states, where subshot noise phase sensitivity is engineered through reducing variance in the measurement axis at the expense of variance in a perpendicular axis according to a Heisenberg uncertainty relation.  Given a set of measurement results, the phase estimation strategy that achieves the degree of quantum enhancement $\xi$ is the method of moments (MOM) strategy, where $ \langle \hat{J}_z \rangle$ is assigned to a phase estimate $\varphi$ \cite{LucaPezze2018}. 
 Assuming the measurement axis is $\hat{J}_z$, the degree of spin-squeezing is expressed in the Wineland spin-squeezing parameter \cite{PhysRevA.46.R6797}
\begin{equation}
\label{eqnspin-squeezing-param}
    \xi = \dfrac{\sqrt{N\text{Var}(\hat{J}_z))}}{|\langle \hat{J}_x \rangle|},
\end{equation}
where $\xi < 1$ implies the state is spin-squeezed along the $\hat{J}_z$ axis, and surpasses the shot noise limit by a factor $\frac{1}{\xi}$.
States with $\dfrac{\sqrt{N}}{\sqrt{F_Q}}<1$ but $\xi >1$ are known as metrologically useful non-Gaussian quantum states \cite{Strobel:2014}, and cannot achieve subshot noise sensitivity with a MOM estimator. We will develop separate frameworks for analysing state preparation error in spin-squeezed and non-Gaussian states.

\begin{figure*}
    \centering
    \includegraphics[width=1\textwidth]{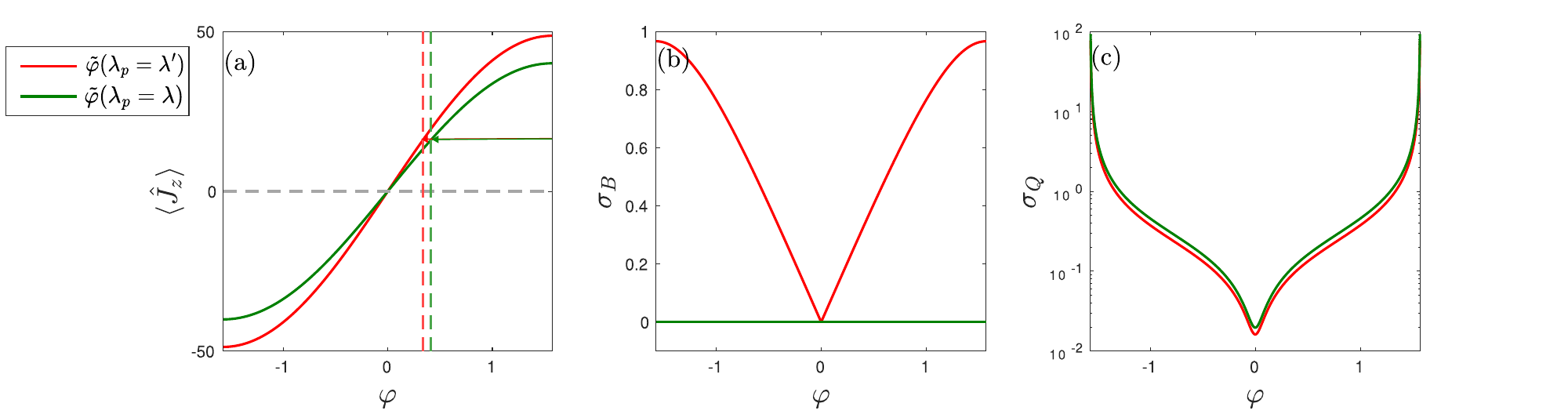}
    \caption{In (a), the average $J_z$ value as a function of the parameter $\varphi$ and the estimator $\tilde{\varphi}$ for biased and unbiased estimators parameterised by $\lambda'=0.01$ and $\lambda=0.02$ respectively for a two-axis twisted spin-squeezed state of $N=100$ (these two values correspond to $\xi = 0.38$ and $\xi = 0.2$ respectively). Given a measured $\langle \hat{J}_z\rangle$, using the incorrect estimator will lead to a biased $\varphi$ value, as demonstrated by the red and green arrows. In (b), the absolute value of the bias using the estimator $\tilde{\varphi}(\lambda_p=\lambda')$ for varying $\varphi$. Where there is no error, as in the estimator parameterised by $\lambda$, there is no bias. In (c). The variance of the biased and unbiased estimators. Variance is minimised at the $\varphi=0$ operating point, and differs for the biased and unbiased estimators.}
    \label{fig:systeestexplain}
\end{figure*}

We now examine how bias can arise, and variance can change under state preparation error for a method of moments estimator and a spin-squeezed state. This estimator involves a particular mapping between $\langle \hat{J}_z\rangle$ and the parameter of interest, with $\langle \hat{J}_z \rangle$ being the average of the measured results $\mathbf{X}$. In the case of spin-squeezed atom interferometry sequences, the unitary $\hat{U}_{\mathbf{\Lambda}}$ from Equation \ref{eqn:genunitary} can be broken up as 
\begin{equation}
    \hat{U}_{\mathbf{\Lambda}} = \hat{U}_\theta \hat{U}_\lambda,
\end{equation}
where $\lambda$ parametrises the strength of spin-squeezing independent of readout, and where $\theta$ parameterises a constant or $\lambda$-dependent $\hat{J}_x$ phase space rotation designed to minimise variance in the measurement axis $\hat{J}_z$. In order to determine the appropriate MOM estimator, we first determine the phase response of the device. Working in the Heisenberg picture, $\hat{J}_z$ evolves as
\begin{equation}
    \label{eqn:Jz expanded}
    \hat{J}_z=-\sin(\varphi)\hat{J}_{x_0}+\cos(\varphi)\left(\hat{J}_{z_0}\cos(\theta)+\hat{J}_{y_0}\sin(\theta)\right), 
\end{equation}
where $\hat{J}_{i_0}$ refers to the operator post the application of $\hat{U}_{\lambda}$ but before the application of $\hat{U}_{\varphi}\hat{U}_\theta$. 
Given that $\ket{\psi_0}$ is a $\hat{J}_x$ eigenstate, we approximate that $\langle \hat{J}_{z_0} \rangle$,$\langle \hat{J}_{y_0}\rangle= 0$ and thus we can write $\langle \hat{J}_z \rangle$ as
\begin{equation}
    \langle \hat{J}_z \rangle = -\langle \hat{J}_{x_0} \rangle\sin(\varphi) \, .
\end{equation}
Informed by this, our estimate of $\varphi$ is then
\begin{align}
\label{eqn:arcsinest}
\varphi^* &= \sin^{-1}\left(-\frac{\langle \hat{J}_z\rangle}{\langle \hat{J}_{x_0}\rangle^\prime}\right) \, ,
\end{align}
where $\langle \hat{J}_{x_0}\rangle^\prime \equiv \langle \hat{J}_{x_0}\rangle$ evaluated at $\lambda_p = \lambda^\prime$, and we restrict the $\varphi$ domain to $-\frac{\pi}{2} \leq \varphi \leq \frac{\pi}{2}$. Expanding Equation \ref{eqn:arcsinest} to linear order around $\langle \hat{J}_z \rangle =0$, we can write
\begin{equation}
    \sigma_B =\varphi \left (1-\dfrac{\langle \hat{J}_{x_0}\rangle'}{\langle \hat{J}_{x_0} \rangle} \right).
    \label{eqn:bias}
\end{equation}
Similarly, we can calculate $\sigma_Q$ via the error propagation equation
\begin{align}
\sigma_Q^2 &= \frac{1}{m}\frac{\mathrm{Var}(\hat{J}_z)}{\left(\partial_\varphi \langle \hat{J}_z\rangle'\right)^2} , \label{sigma_Q_simp}
\end{align}
where the numerator is based on the measured variance in $\langle \hat{J}_z\rangle$ and the denominator is evaluated at the assumed state preparation value $\lambda^\prime$. Using Equation \ref{eqn:Jz expanded}, this becomes
\begin{align}
\sigma_Q^2 &=  \dfrac{1}{m|\langle \hat{J}_{x_0}\rangle'|^2}\left(\cos^2(\theta)\text{Var}(\hat{J}_{z_0}) + \sin^2(\theta)\text{Var}(\hat{J}_{y_0}) \right. \nonumber \\
&\quad + \left.\frac{1}{2}\sin(2\theta)\overline{\text{Covar}}(\hat{J}_{z_0},\hat{J}_{y_0}) \right),
\label{eqn:chap5var}
\end{align} 
where \ref{eqn:chap5var} is derived in Appendix \ref{app:bigeqder} A, and $\bar{\text{Covar}}(a,b) =\text{Covar}(a,b) + \text{Covar}(b,a) $. Figure \ref{fig:systeestexplain} (a) shows $\langle \hat{J}_z\rangle$ for two different values of $\lambda$, corresponding to $\xi = 0.20$ and $\xi = 0.38$, leading to different phase responses. Assuming one value of $\lambda$, while in reality, the other is true, leads to a miss-estimate of $\varphi$. Processing $\langle\hat{J}_z\rangle$ with the wrong estimator leads to a biased estimate of $\varphi$, with the red and green arrows in (a) demonstrating a biased estimate given a $\langle \hat{J}_z \rangle$ measurement at a particular operating point. In (b) and (c) we see $\sigma_B$ and $\sigma_Q$ respectively. In this case, miss-characterising the state leads to a bias that is approximately linear in $\varphi$.  The variance of the estimated parameter around the mean also changes relative to the unbiased estimator, that is, $Q(\lambda) \neq Q(\lambda')$, in Figure \ref{fig:systeestexplain} (c). In this particular example, the noise in the measurement data $\langle \hat{J}_z\rangle$ is not changed through state preparation error; the change in $Q$ is due purely to the use of a different estimator affecting the denominator of Equation \ref{sigma_Q_simp}. We will also see cases in Section \ref{OATsec} where the data $\mathbf{X}$ are affected by a poorly characterised $\lambda$.

\subsection{Spin-squeezed OAT vs TAT}\label{OATsec}
We consider two spin-squeezing methods to study spin-squeezed states under state preparation error, namely two-axis twisting (TAT) \cite{Kitagawa1993, PhysRevA.80.012318} and one-axis twisting (OAT) \cite{Esteve2008, PhysRevLett.86.4431, Haine:2014, Schleier-Smith:2010, Riedel:2010}. TAT has yet to be experimentally realised, but provides the ``simplest" example of an entangled state preparation sequence, given that its optimal $\hat{J}_x$ rotation parameter is independent of $\lambda$. It has been shown that TAT dynamics can be produced in cavity-mediated atom-atom interactions \cite{Luo:2024}, and can produce a state similar to that prepared by quantum-nondemolition measurement (QND) squeezing \cite{Kuzmich2000,Schleier-Smith:2010, Haine:2015, Hosten:2016, Kritsotakis:2021, Greve:2022}. For an initial $\hat{J}_x$ eigenstate, the TAT Hamiltonian and state preparation sequence are 
\begin{align}
    \hat{H}_{\text{TAT}} &= -\hbar\chi(\hat{J}_z\hat{J}_y+\hat{J}_y\hat{J}_z),\\
    \label{eqn:TATspinsqueezeunitary}
    \hat{U}_{\text{TAT}} &= \exp{\left(\frac{-i \pi\hat{J}_x}{2}\right)}\exp{\left(i\lambda(\hat{J}_z\hat{J}_y+\hat{J}_y\hat{J}_z)\right)},
\end{align}
where $\lambda = \chi t$.
OAT has also been experimentally demonstrated \cite{Gross:2010, Riedel:2010, Leroux:2010}. The OAT Hamiltonian is given by
\begin{equation}
    \hat{H}_{\text{OAT}} = - \hbar \chi \hat{J}_z^2,
\end{equation}
 and the OAT state preparation sequence is given by
\begin{equation}
    \label{eqn:OATspinsqueezedunitary}
    \hat{U}_\text{{OAT}} = \exp{\left(-i\theta (\lambda') \hat{J}_x\right)}\exp{\left(i\lambda\hat{J}_z^2\right)},
\end{equation}
where the optimal $\hat{J}_x$ rotation $\theta$ is dependent on $\lambda$, and thus a suboptimal rotation will be applied if $\lambda' \neq \lambda$.

\begin{figure*}[t]
    \centering
    \includegraphics[width=1\textwidth]{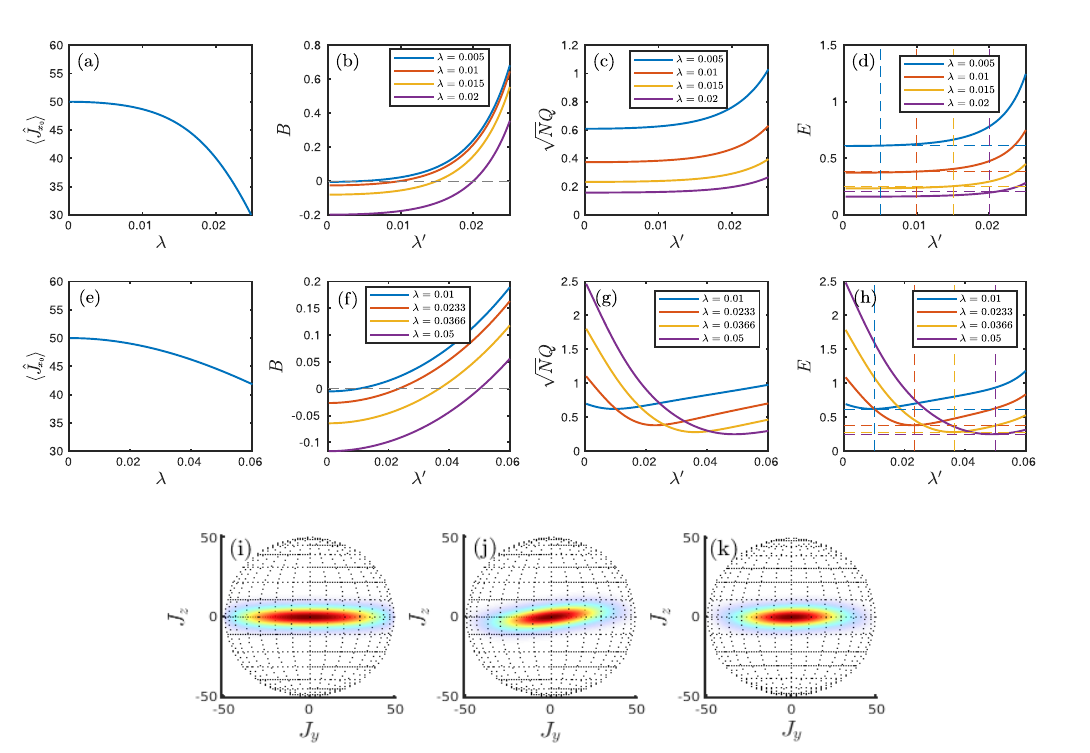}
    \caption{In (a) and (e), plots of $\langle \hat{J}_{x_0} \rangle$ as a function of the state preparation parameter $\lambda$ for OAT and TAT, respectively. In (b) and (f), the bias coefficient $B$ for TAT and OAT for an assumed state preparation parameter $\lambda'$ for a set of states parameterised by varying $\lambda$. In (c), (d) and (g), (h), the single shot variance $Q$, and the lower bound on error $E$ for OAT and TAT respectively given the same $\lambda$ and $\lambda'$ values as in (b) and (f). $\lambda \neq \lambda'$ leads to a biased estimator $\tilde{\varphi}(\lambda_p=\lambda')$, leading to a biased parameter estimate, with the magnitude of the bias dependent on the discrepancy between $\langle \hat{J}_{x_0} \rangle$ and $\langle \hat{J_{x_0}}\rangle '$ for both OAT and TAT, as per Equation \ref{eqn:bias}. Biased estimators lead to superior $Q$ and $E$ in (c) and (d) for TAT for $\lambda' < \lambda $, whilst the unbiased estimator is the optimal estimator for OAT in (g) and (h). This distinction is due to the state preparation parameter dependence of the $\hat{J}_x$ rotation for OAT, but not for TAT. In (i) we see a $\lambda = 0.02$ TAT state aligned to the measurement axis $\hat{J}_z$ for all $\lambda'$ values, whereas in (j) and (k), we see differing alignment to the $\hat{J}_z$ axis for $\lambda= 0.034, \lambda' =0.014$, and $\lambda=\lambda'=0.034$. }
    \label{fig:figure1_results}
\end{figure*}

We simulate the TAT and OAT interferometry sequence according to Equations \ref{eqn:OATspinsqueezedunitary} and $\ref{eqn:TATspinsqueezeunitary}$ for $N=100$ atoms in an initial $\hat{J}_x$ eigenstate given a sample number $m = 10^4$, a phase $\varphi=\frac{Q}{\sqrt{m}}$ encoded anti-clockwise around the $\hat{J}_y$ axis, and a range of actual and assumed $\lambda$ and $\lambda'$ values. For each $\lambda$, $\lambda'$ pair, we calculate $B$, $Q$, and $E$ according to Equations \ref{eqn:bias}, \ref{eqn:chap5var}, and \ref{eqn:errormetric} respectively. In the case of both OAT and TAT, we find that $\lambda' \neq \lambda$ leads to a biased estimate of the encoded phase as $\tilde{\varphi}$ is parameterised by $\lambda_p=\lambda'$ instead of $\lambda_p=\lambda$. In both cases, $\lambda' <\lambda$ leads to negative bias and $\lambda' > \lambda$ leads to positive bias, as can be seen in Figure \ref{fig:figure1_results} (b) and (f). The magnitude and sign of the bias depend on the discrepancy between the assumed and actual $\langle \hat{J}_{x_0} \rangle$ values, as per Equation \ref{eqn:bias}, with the dependency of $\langle \hat{J}_{x_0} \rangle$ for both TAT and OAT displayed in Figure \ref{fig:figure1_results} (a) and (e) respectively. For both OAT and TAT in the spin-squeezing regime, the $\langle \hat{J}_{x_0}\rangle$ value begins to decrease at an increasing rate as $\lambda$ increases. This results in the slope of the graph of $B$ increasing for both OAT and TAT with respect to the assumed state preparation parameter.

The plots of $E$ for OAT and TAT are qualitatively distinct. In the case of TAT in Figure \ref{fig:figure1_results} (d), we find that $E$ increases uniformly as the assumed state preparation value $\lambda'$ increases, with $E$ for $\lambda' < \lambda$ lower than for $\lambda'>\lambda$. This has a surprising implication: that the ``wrong" estimator may be capable of improving the precision of an atom interferometry experiment. Whilst $E$ is a lower bound on the achievable error, given the small contribution of the $\varphi$ term in Equation \ref{eqn:chap5var} and the neighbouring plot of $Q$ in (c), we expect that $E$ will be close to the actual error achievable for some $m$ and $\varphi$ values. We will investigate this in the next subsection. The decrease in $Q$ comes because the variance of the parameter estimate is reduced as a result of underestimating $\langle \hat{J}_{x_0}\rangle$ according to Equation \ref{eqn:chap5var}. The increased rate of $Q$ increase for $\lambda'>\lambda$ owes to the increased rate of decrease in $\langle \hat{J}_{x_0}\rangle$ in the same expression. For OAT, we find that $E$ is optimal when $\lambda'=\lambda$. This is because state preparation error in OAT leads to the ``wrong" $\hat{U}_{\theta}$ being applied, meaning that the axis of maximal spin-squeezing is not along the measurement axis $\hat{J}_z$ - the data $\mathbf{X}$ produced are suboptimal. As a result, greater single shot variance ensues, as can be seen in (g). We note that $Q$ climbs much more steeply for $\lambda' < \lambda $ than for $\lambda' > \lambda$. This owes to the fact that the optimal $\theta$ angle changes much more quickly for low $\lambda$ values than for higher $\lambda$ values.

The qualitative distinction between $E$ for OAT and TAT can be understood by considering the Q-sphere phase space representations of TAT and OAT states before parameter encoding in Figure \ref{fig:figure1_results} (i), (j), and (k). In (i), we see a TAT quantum state for $\lambda = 0.02$ for all $\lambda ' $ values. As the $\hat{J}_x$ rotation angle does not depend on $\lambda$, the axis of maximal spin squeezing is always aligned to the $\hat{J}_z$ axis. In (j), we see an OAT state with $\lambda = 0.034$ prepared with an assumed value $\lambda'=0.014$, which, in contrast to (i), results in the axis of maximal spin squeezing not aligning to the readout axis $\hat{J}_z$ as $\theta$ is dependent on $\lambda$. The corresponding OAT state for $\lambda=\lambda'=0.034$ is displayed in (k), and is aligned to the $\hat{J}_z$ readout axis.

The quantum state that is most robust to state preparation error in Figure \ref{fig:figure1_results} is the TAT state for small $\lambda$ ($\lambda=0.005$). This is not only because $\theta$ is independent of $\lambda$, but also because its $\langle \hat{J}_{x_0} \rangle$ moment changes negligibly with $\lambda$ around this region. By Equations \ref{eqn:bias} and \ref{eqn:chap5var}, this means bias will be small, and $Q$ will nearly be equal to $Q(\Delta\lambda=0)$. In general, thsee equations suggest spin-squeezed quantum states around which $\partial_{\lambda}\langle \hat{J}_{x_0}\rangle =0$ (and where $\Delta\lambda \neq 0$ doesn't influence later state preparation parameters) will be robust to state preparation error.

\subsection{Bias-variance tradeoff}
In Figure \ref{fig:figure1_results} (d), we see for $\lambda' < \lambda$ that $E$ decreases below the error for a TAT interferometry sequence free from state preparation error (for $\Delta \lambda =0$, $E$ is saturated). Using an estimator $\tilde{\varphi}(\lambda_p=\lambda')$ rather than $\tilde{\varphi}(\lambda_p=\lambda)$ led to a decreasing lower bound on the state preparation error. Although this did not demonstrate that a biased estimator would necessarily reduce MSE, it indicated that such a scenario may arise for TAT, since we expect $E$ to be close to actual MSE. To investigate how using a biased estimator may decrease MSE, we fix the parameter of interest at $\varphi=0.001$, and plot the ratio of the lowest mean squared error $\text{MSE}_\text{opt}$ to MSE$(\lambda_p=\lambda)$ given a set of estimators $\tilde{\varphi}(\lambda_p)$ with $\lambda_p$ spanning from $0$ to $\lambda$, over a range of state preparation parameters and sample sizes.

\begin{figure}
    \centering
    \includegraphics[width=1\linewidth]{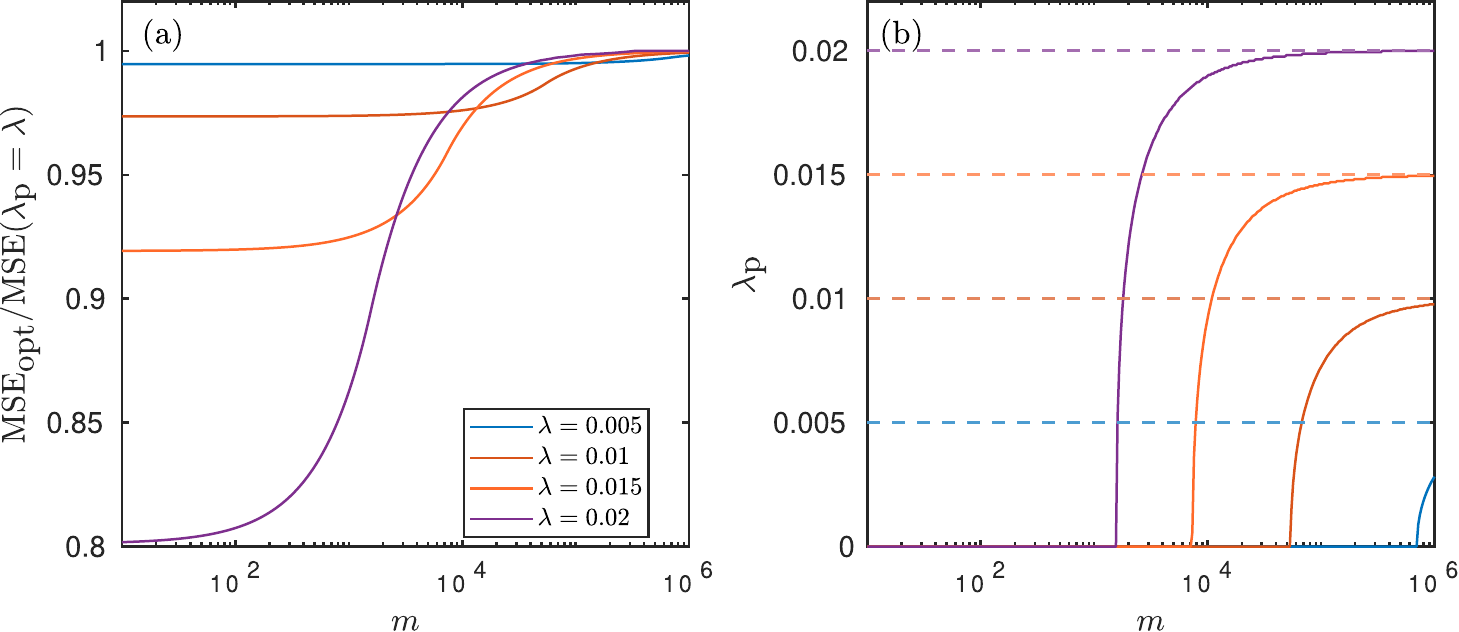}
    \caption{(a): the ratio of the optimal MSE for a biased estimator divided by the MSE of the unbiased estimator, for a range of $\lambda$ and shot numbers $m$ for two-axis-twisting. The $\lambda_p$ values parameterising the biased estimator $\tilde{\varphi}(\lambda_p)$ are $0 \leq  \lambda_p < \lambda$. Biased estimators with lower noise are able to achieve increased precision over unbiased estimators, with this advantage dissipating for larger $m$ due to bias staying constant and variance decreasing as $\frac{1}{\sqrt{m}}$. (b): the $\lambda_p$ value that optimises the MSE for each state preparation value with the $\lambda$ values displayed as $y$-axis asymptotes. For larger $\lambda$, bias more heavily counteracts the decreased variance from the biased estimator, meaning that the optimal $\lambda_p$ gets closer to $\lambda$ for smaller $m$. }
    \label{fig:tatbiasvar}
\end{figure}

In Figure \ref{fig:tatbiasvar} (a), we see that for every $\lambda$, the unbiased estimator never produces the least error - the optimal ratio between the best biased MSE and the unbiased MSE is always less than 1. We thus see that atom interferometry with TAT states would be amenable to the bias variance trade-off, where a biased estimator decreases overall error relative to an unbiased estimator because of its lower variance.

The bias variance trade-off becomes more useful for higher $\lambda$ values. This owes to the fact that $\langle\hat{J}_{x_0}\rangle$ is smaller for higher $\lambda$ as shown in Figure \ref{fig:figure1_results} (a), meaning the denominator in Equation \ref{eqn:chap5var} can be artificially increased by picking an estimator that ``overestimates" $\langle \hat{J}_{x_0} \rangle$. We note that the larger $m$ is, the lower the achievable bias variance trade-off. This is because as $m$ increases, the non-zero bias term from the biased estimator stays constant, whilst $\sigma_Q$ decreases as $\frac{1}{\sqrt{m}}$ - the decrease in the variance possible from choosing a biased estimator is counteracted by the bias. In Figure \ref{fig:tatbiasvar} (b), we see the $\lambda_p$ value parameterising the MSE-optimising estimator for a range of $\lambda$ and $m$ values. The $m$ value at which the optimal $\lambda_p \approx \lambda$ decreases for more highly squeezed states. This is because more highly squeezed states are more susceptible to the biasing effects of state preparation error. As per Figure \ref{fig:figure1_results} (b), this bias is reduced by having $\lambda_p$ closer to $\lambda$, where we interpret $\lambda'$ as being $\lambda_p$ in this subsection.
The above results suggest that biased estimators may lead to lower measurement error if TAT can be realised and manipulated in an atom interferometry experiment. For other state preparation schemes where the measurement results $\mathbf{X}$ are not affected by $\lambda'$, it is likely that similar bias variance trade-offs exist.

\subsection{Varying atomic ensemble size}
We now investigate the effect of state preparation error on atomic ensembles large enough to produce useful measurements ($N>10^4$). Numerically simulating an interferometry sequence with this number of atoms is intractable, due to the size of the Hilbert space. To probe the sensitivity of larger number of atoms under systematic state preparation error, we use the analytic expressions for the moments of OAT states in Appendix \ref{app:oatmom}, combined with Equations \ref{eqn:bias} and \ref{eqn:chap5var}.
We assume that an experimenter seeks to produce maximally spin-squeezed states. The maximally spin-squeezed OAT state is reached at a $\lambda$ value of \cite{Kitagawa1993}
\begin{equation}
   \lambda(N) =\frac{24^{1 / 6}}{2^{1 / 3} N^{2 / 3}},
\end{equation}
where the above expression is derived in the limit of small $\lambda$ which is valid so long as $N>>1$. We define $\Delta \lambda_{\text{crit}}$ as the value of $|\lambda'-\lambda|$ such that $E$ passes a certain threshold (in the case of our results, $\Delta \lambda <0$ when these thresholds are reached). This places a bound on the maximum sensitivity achievable with a $N$-atom OAT ensemble under state preparation error at the $\lambda'$ where this threshold is passed for all $\varphi$ operating points.

We first consider a set of thresholds relative to the quantum noise of an unbiased state preparation scheme - $E(\Delta\lambda=0)$. We also consider a set of thresholds relative to the amount of error achievable with an unentangled state, denoted $E_\text{un}$. 
\begin{figure*}[t!]
    \centering
    \includegraphics[width=1\textwidth]{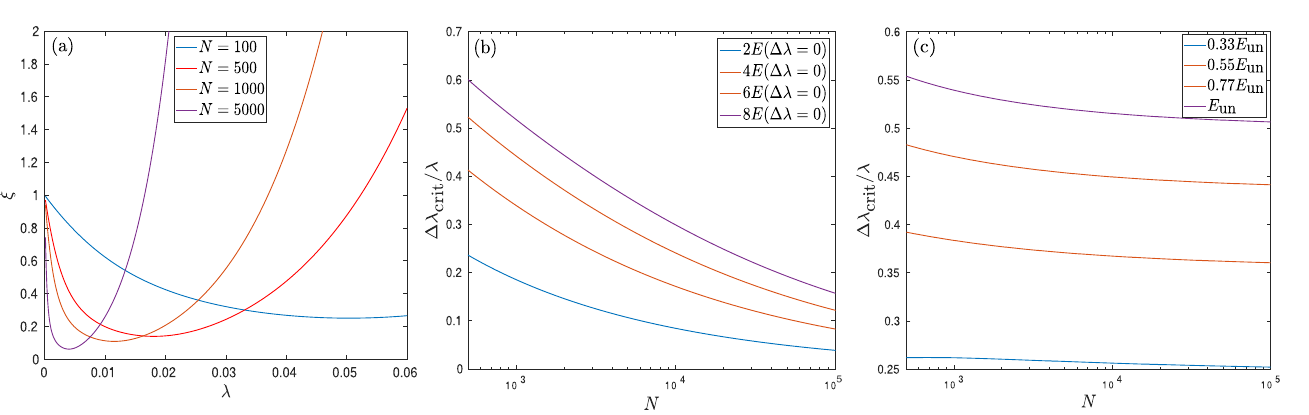}
    \caption{(a): Amount of spin-squeezing as a function of the state preparation parameter $\lambda$ for varying one-axis-twisted atomic ensemble sizes $N$. (b) and (c):  $\frac{|\lambda'-\lambda|}{\lambda}$ values such that error is at least $2 - 8$, times $E(\Delta \lambda=0)$, and $0.33-1$ times the shot-noise limit $E_\text{un}$ respectively. In (b), the larger the ensemble size, the less state preparation error required to reduce the efficacy of spin-squeezing by a constant factor relative to the sensitivity of the state produced. This is due to greater spin-squeezing for larger atomic number, with more spin-squeezed states being more sensitive to suboptimal $J_x$ rotation angles.  However, relative to an unentangled state in (c), we see that the robustness of OAT states for increasing atomic ensemble size is nearly constant. }  
    \label{fig:numberdeltacrit}
\end{figure*}

We plot $\frac{\Delta \lambda_{\text{crit}}}{\lambda}$ for a range of thresholds as a function of $N$. In Figure \ref{fig:numberdeltacrit} (b), we see $\frac{\Delta \lambda_{\text{crit}}}{\lambda}$ for $E(\Delta \lambda )/E(\Delta \lambda=0 ) = 2$, $4$, $6$, and $8$. For each of these thresholds, less percentage error in $\lambda$ is required to cause an equivalent error factor as $N$ increases. This is because as $N$ increases, so does the maximal spin-squeezing one can generate, as shown in Figure \ref{fig:numberdeltacrit} (a). More squeezed states are more susceptible to suboptimal $\hat{J}_x$ rotation angles, which means a lower percentage of systematic error is required to decrease the precision by a particular factor. In (c), we see that roughly equivalent magnitudes of state preparation error at varying $N$ values lead to the same precision decrease relative to constant fractions of shot noise. This suggests that squeezing under state preparation does not get much worse for large $N$ relative to the precision achievable with unentangled ensembles of atoms.
 
The values in Figure \ref{fig:numberdeltacrit} (b) and (c) suggest that entangled state preparation error is a significant hindrance to precision in quantum-enhanced atom interferometry. Given the multitude of ways of realising OAT,  we will not attempt to analyse state preparation error in the context of a specific experimental configuration. However, our results provide a useful starting point for an experimenter wishing to do so. More generally, they offer a schematic for how to calculate error in parameter resolutions under a particular state-generation Hamiltonian for atomic ensembles large enough to produce metrologically useful measurements.

\section{Systematic state preparation error in non-Gaussian states}
To systematically surpass the shot-noise limit with a non-Gaussian state, one cannot use a method of moments estimator. An estimator that will saturate the precision limit for a particular basis is the maximum likelihood estimator. The data $\mathbf{X}$ operated on by this estimator are the entire distribution of measurement results, rather than a moment of the results. The $\varphi^*$ value is determined by calculating the $\varphi$ value that maximises an objective function for the set of experimental results, known as a likelihood function, denoted $\mathcal{L}$. Maximum likelihood estimation in quantum-enhanced atom interferometry is performed according to the equation
\begin{equation}
    \label{eqn:likelihoodstateprep}
  \text{argmax}_\varphi \mathcal{L}_{\lambda_p}(\mathbf{X}|\varphi) = \text{argmax}_\varphi \sum^{m}_{i=1}\text{log}(|\braket{X_i}{\psi_{\varphi,\lambda_p}}|^2 ),
\end{equation}
where the data are produced by the actual state preparation parameter vector $\mathbf{\Lambda}$, and where $\ket{\psi}$ is parametrised by both the estimator state preparation parameter $\lambda_p$ and the parameter $\varphi$ that we wish to optimise over. Under state preparation error where $\lambda_p=\lambda'$ for $\lambda' \neq \lambda$, a biased likelihood function is optimised, a scenario termed maximum likelihood estimation under model misspecification. Using a biased likelihood function for parameter estimation will mostly lead to biased estimates of the parameter of interest.

For an unbiased estimator in the optimal basis, the single shot variance $Q$ aligns with the Quantum Cramér-Rao Bound (QCRB). However, under model misspecification, the single-shot variance of the estimate will not necessarily agree with the QCRB and is given by
\begin{equation}
    Q^2=\dfrac{\text{Var}(Z_{\lambda_p}(X, \varphi^*))}{\mathbf{E}[\partial_{\varphi}Z_{\lambda_p}(X, \varphi^*)]^2},
\end{equation}
where $Z_{\lambda_p}(X, \varphi) = \frac{\partial}{\partial \varphi} \log(|\braket{X}{\psi_{\varphi, \lambda_p}}|^2)$ and where $X$ denotes the random variable corresponding to measurement outputs. 
By expanding the unbiased likelihood function to second order in the state preparation degree of freedom $\lambda$, and the parameter of interest $\varphi,$ we find in Appendix \ref{app:MLE_S} that the bias for a small error in $\lambda$ is
\begin{equation}
    \label{eqn:Smetric}
    \sigma_B = -\dfrac{F_{\varphi \lambda}}{ F_{\varphi \varphi}}(\lambda' - \lambda),
\end{equation}
where $F_{\varphi \lambda}$ and $F_{\varphi \varphi}$ are the off-diagonal and second diagonal terms of the Fisher covariance matrix evaluated at the actual $\lambda$ and $\varphi$ values, the entries of a Fisher information matrix given by
\begin{equation}
    F_{ij}= \sum_{m} \dfrac{\partial_i P_m\partial_j P_m}{P_m}.
\end{equation}
This suggests that the ratio $\frac{F_{\varphi\lambda}}{F_{\varphi\varphi}}$ encapsulates the local susceptibility of a quantum state to state preparation error in a particular measurement basis. If this ratio is near $0$ for a continuous range of $\lambda$ values, the state should be immune from bias under state preparation error. Furthermore, assuming that all other state preparation parameters are independent of $\lambda$, the misspecified model will remain constant under state preparation error, leading to $Q$ also remaining constant.

\subsection{Non-Gaussian OAT and TNT}
 We investigate OAT and TNT non-Gaussian states under state preparation error using Monte Carlo techniques for small state preparation errors, and compare them to the Taylor expansion-based estimate. The TNT Hamiltonian is defined by 
\begin{equation}
    \label{eqn:TNTham}
    \hat{H}_{\text{TNT}}=\hbar(\chi\hat{J}_z^2 +\Omega \hat{J}_x)
\end{equation}
where $\Omega = \frac{\chi N}{2}$ \cite{Muessel:2015, PhysRevA.63.055601, Mirkhalaf2018}.
Non-Gaussian OAT states have yet to be produced, whilst non-Gaussian TNT states have been produced but have yet to yield subshot noise precision in an atom interferometry experiment \cite{Strobel:2014}. The QFI for a non-Gaussian OAT is maximised without a $\hat{J}_x$ rotation, and thus the state preparation sequence is
\begin{equation}
    \label{eqn:OATprotocolnonGaussian}
    \hat{U}_{\text{OAT}}=e^{i\lambda\hat{J}_z^2}.
\end{equation}
 In the case of TNT, the QFI can be increased by applying an appropriate $\hat{J}_x$ rotation just before parameter encoding. Whilst the optimal rotation is dependent on the state preparation parameter $\lambda$, highly metrologically useful states can be obtained with $\theta=-1$, with the upside of a system that is simpler to experimentally implement and analyse. This results in a TNT state preparation sequence of 
\begin{equation}
    \label{eqn:TNTnonGaussianprotocol}
    \hat{U}_{\text{TNT}}=e^{i\hat{J}_x}e^{-i(\lambda \hat{J}_z^2+\delta \hat{J}_x)},
\end{equation}
where $\lambda=\chi t$ and $\delta = \Omega t$, with the optimal value of $\delta$ being $\delta = \frac{\lambda N}{2}$.
For both OAT and TNT, we seek to perform an optimal readout, i.e. the readout that saturates the QCRB. The optimal readout basis for OAT and TNT non-Gaussian states is $\hat{J}_x$ \cite{SamuelP.Nolan2017, Mirkhalaf2018, Haine2018}. A readout in this basis can be realised through a $\pi/2$ rotation clockwise about the $\hat{J}_y$-axis, followed by a standard number-difference measurement. Thus, we attempt to find the value $\varphi$ that maximises the value of the log-likelihood function
\begin{equation}
    \label{eqn:likelihoodisngle}
  \mathcal{L}_{\lambda_p}(\mathbf{X}|\varphi) =  \sum^{i=m}_{i=1}\text{log}\left(|\bra{X_i}e^{i(\frac{\pi}{2}-\varphi\hat{J}_y)}\hat{U}_{\text{OAT/TAT}} \ket{\psi_0}|^2\right),
\end{equation}
for OAT and TNT respectively. 

We simulate the generation, readout, and estimation of a parameter encoded onto the quantum states through Monte-Carlo sampling. We perform the simulation for states in both the spin-squeezed and non-Gaussian regimes, with the non-Gaussian regime starting for OAT at $\lambda \approx 0.1$, and for TNT at $\lambda \approx 0.045$. For OAT, we consider $\lambda' = \lambda  \pm0.0025$ for $45$  $\lambda$ values between $0.01$ and $0.17$, and for TNT, we consider $\lambda' = \lambda \pm 0.0015$ for $45$ $\lambda$ values between $0.01$ and $0.11$. For each $\lambda,\lambda'$ pair, we perform a maximum likelihood estimate parameterised by $\lambda_p =\lambda'$ given $10^5$ measurements of the prepared quantum states, and repeat these $5$ and $25$ times for each $\lambda, \lambda'$ pair for OAT and TNT respectively. We calculate the bias, and the ratio of the quantum noise $Q$ to the QCRB, comparing the biased estimate to that derived from Equation \ref{eqn:Smetric}. Given that bias is not proportional to the parameter of interest for an MLE estimator, a similarly useful lower bound to $E$ for spin-squeezed states cannot be devised, and the parameter of interest is fixed at $\varphi=0.02$ for the simulations.

\begin{figure*}
    \centering
    \includegraphics[width=0.75\textwidth]{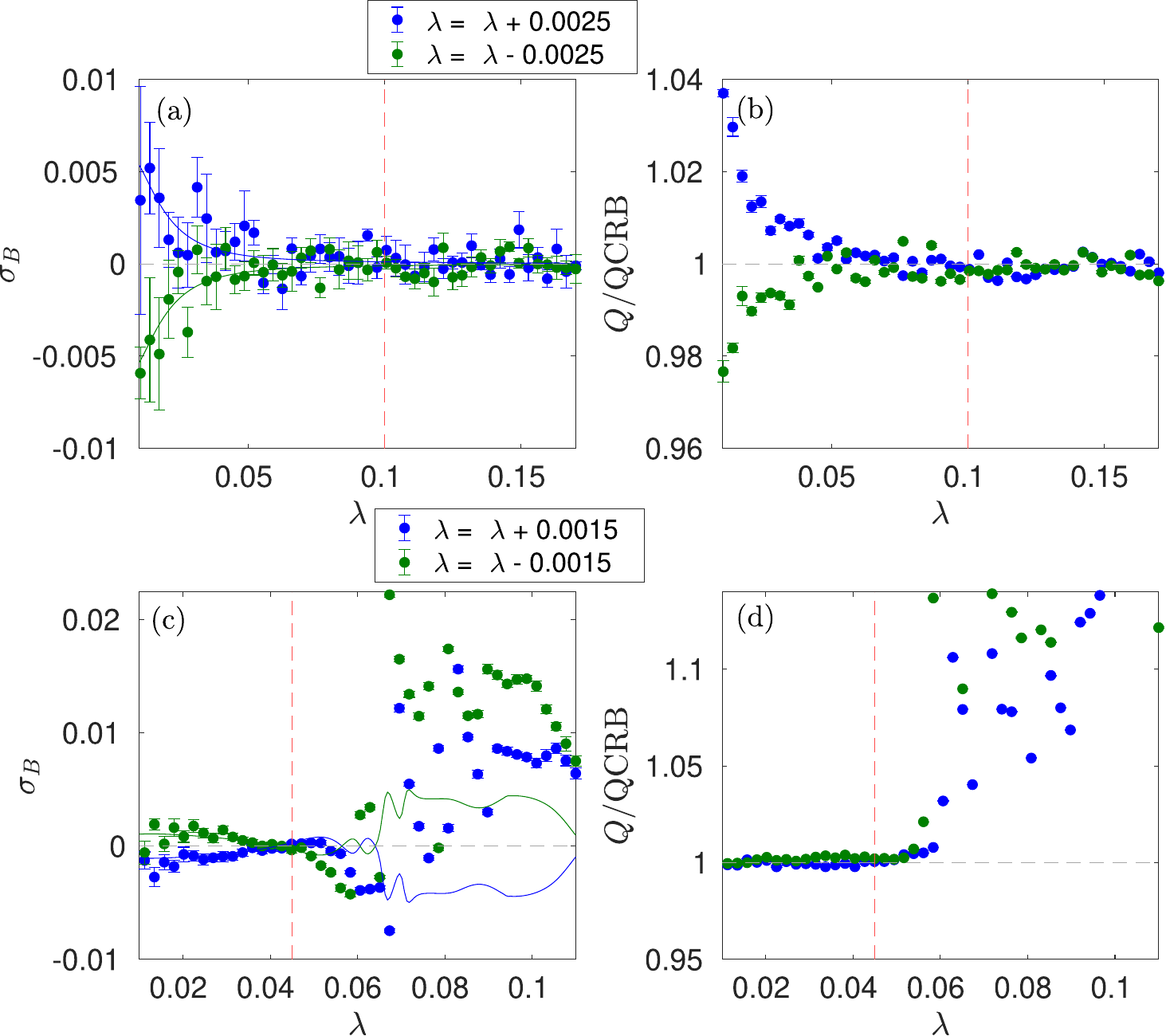}
    \caption{In (a) and (c) the bias for OAT and TNT states in their respective Gaussian and non-Gaussian regimes for state preparation parameter $\lambda$ and assumed state preparation parameter $\lambda'$ under misspecified maximum likelihood estimation, compared with the bias predicted by Equation \ref{eqn:Smetric}. In (b) and (d), the ratio between single-shot error $Q$ and the Quantum Cramér-Rao bound for OAT and TNT respectively under the same $\lambda,\lambda'$ values as in (a) and (c). OAT in its non-Gaussian regime is robust to state preparation error as can be seen in both (b) and (d), and the metric Equation \ref{eqn:Smetric} is able to predict the bias in both Gaussian and non-Gaussian regimes. TNT in its non-Gaussian regime is not robust to state preparation error, as it is too sensitive to its state preparation parameter in the non-Gaussian for Equation \ref{eqn:Smetric} to be accurate for even the small $\Delta \lambda$ chosen.}
        \label{fig:fig_non_gauss}
\end{figure*}

In the case of OAT in Figure \ref{fig:fig_non_gauss} (a), we see that the covariance matrix ratio in Equation \ref{eqn:Smetric} reliably predicts the bias in both the Gaussian and non-Gaussian regimes, where the start of the non-Gaussian regime is denoted by the dotted red line. In (b), we see that under small state preparation error in the non-Gaussian regime, OAT is essentially immune to state preparation error - $Q/\text{QCRB}$ remains constant. This is because the likelihood function being optimised changes negligibly from the unbiased likelihood function despite $\lambda_p \neq \lambda$, which can be inferred from the small $\frac{F_{\varphi\lambda}}{F_{\varphi\varphi}}$. In the case of TNT, we see that the approximation correctly predicts bias in the Gaussian regime, but is inaccurate past early in the non-Gaussian regime. This can be attributed to higher-order derivatives being required to characterise the likelihood function even at this small amount of state preparation error - the metric of Equation \ref{eqn:Smetric} is of limited use in the absence of likelihood functions that change sufficiently slowly with respect to the state preparation parameter. We see that in contrast to OAT, $Q$ in the case of non-Gaussian TNT states strays significantly from the QCRB even for small state preparation error. Currently, the only non-Gaussian atomic states that have been generated with a significant ($N >10^3$) atomic ensemble are non-Gaussian TNT state. Our results indicate that state preparation error is likely to degrade the overall precision of metrology with non-Gaussian TNT states.

\subsection{Comparing OAT estimation strategies}
OAT's robustness to state preparation error under a maximum likelihood estimator raises an interesting proposition - that under a drifting state preparation parameter, such as in the case of phase diffusion, one would be better off dispensing with a method of moments of estimator in favour of a maximum likelihood estimator, even in the spin-squeezing regime. To investigate this, we compare the $\sigma_B$ value for a MOM estimator and an MLE estimator for constant $\Delta \lambda$ in the absence of any $\hat{J}_x$ rotation. We calculate $\sigma_B$ for the MLE estimator through the first order Taylor expansion in Equation \ref{eqn:Smetric}, as this approximation is reliable in the case of OAT for larger $\Delta \lambda$, and calculate $\sigma_B$ for the spin-squeezed state with Equation \ref{eqn:bias}. For the OAT states where an MLE estimator is used, the measurement occurs in the $\hat{J}_x$ basis as opposed to the $\hat{J}_z$ basis.

\begin{figure}
    \centering
    \includegraphics[width=1\linewidth]{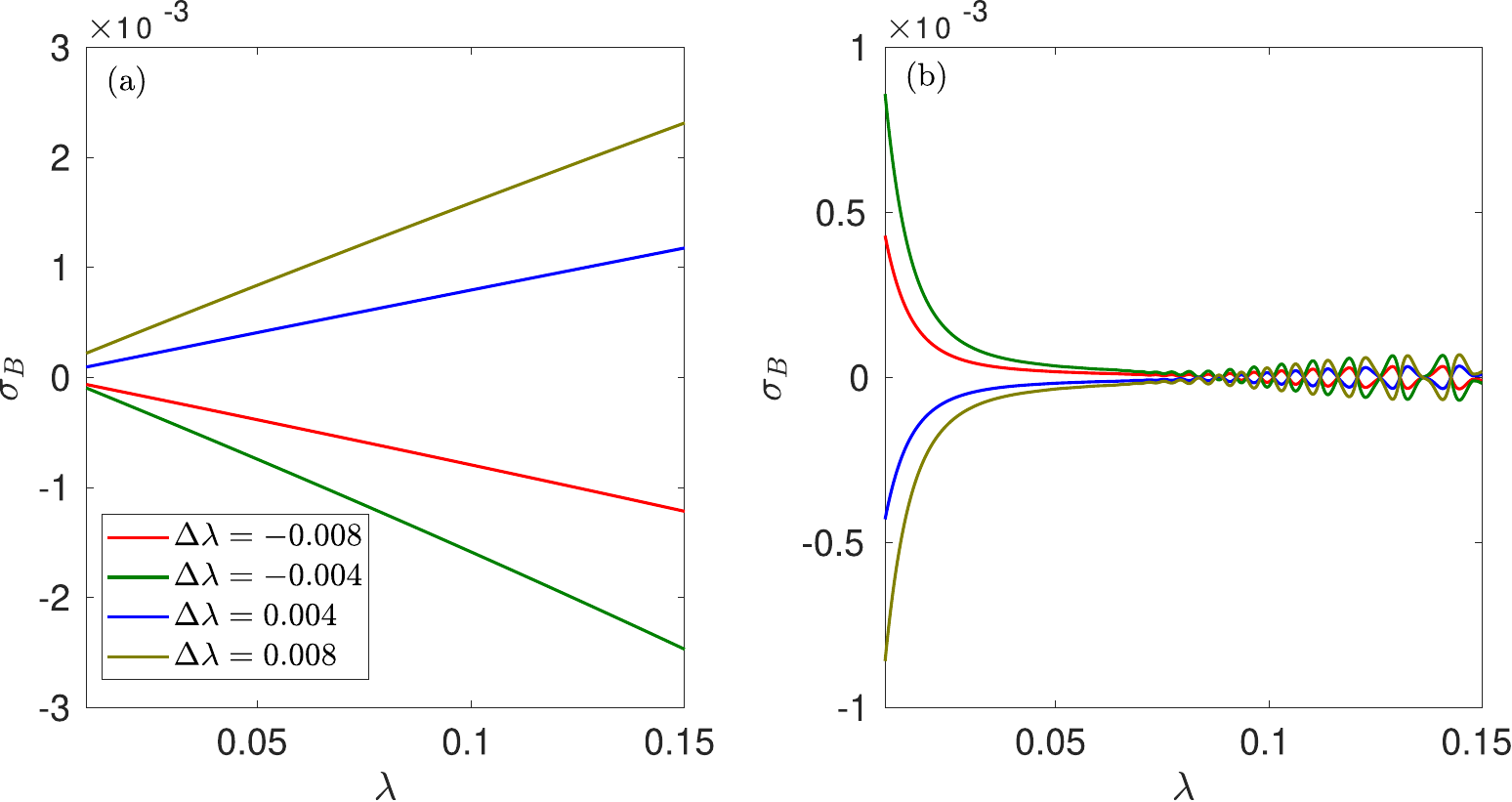}    
    \caption{In (a) and (b) the bias for OAT spin-squeezed states as a function of $\lambda$ for varying $\Delta \lambda$ for MOM and MLE estimators given $J_z$ and $J_x$ bases respectively. For $\lambda>0.015$, bias can be significantly reduced through measuring in the $\hat{J}_x$ basis and processing the measurement data with an MLE estimator, rather than measuring in $\hat{J}_z$ and using an MOM estimtor.}
    \label{fig:OAT_final_fig}
\end{figure}

In Figure \ref{fig:OAT_final_fig} (a) and (b), we see the $\sigma_B$ value for an OAT state for varying $\lambda$ and constant $\Delta \lambda$ given a fixed $\varphi$ value of $0.02$, for an MOM and MLE estimator respectively. Whilst $\sigma_B$ is smaller in the case of the MOM estimator for $\lambda<0.015$, the MOM estimator dramatically reduces the total bias for larger $\lambda$ values. This suggests that under considerable phase diffusion for larger $\lambda$ values, it is wise to use the MLE estimator instead of the canonical MOM estimator. 

\subsection{Mitigating bias in quantum-enhanced atom interferometry}
In both spin-squeezed and non-Gaussian states, we see that systematic state preparation error mostly leads to bias in parameter estimation. Whilst we show that there exist classes of spin-squeezed and non-Gaussian quantum states that are robust to bias from state preparation error, it may not always be possible to manufacture such a state. We present a solution that keeps the estimate of the encoded parameter $\varphi$ unbiased - we discard our assumed value of the state preparation degree of freedom $\lambda'$, and estimate both $\lambda$, and $\varphi$ according to two-parameter maximum likelihood estimation.

For a two-parameter maximum likelihood estimate of a state preparation parameter $\lambda$ and a relative phase $\varphi$, the sensitivity of an estimate of $\varphi$ is given by one on the square root of the upper-diagonal term of the Fisher information matrix, 
\begin{equation}
    \label{eqn:multiparamvarphii}
    Q = \sqrt{ \dfrac{F_{\lambda\lambda}}{F_{\varphi\varphi}F_{\lambda\lambda}-F_{\varphi\lambda}^2}}.
\end{equation}
\sloppy
In a non-diagonal Fisher covariance matrix, $Q_{\text{two-param}}>Q_{\text{one-param}}$ - estimating both parameters at once leads to more error in $\varphi$, than using $\lambda_p = \lambda$, and estimating $\varphi$. However, under state preparation error that induces bias, overall error may be lower under two parameter estimation for sufficient sample size. 

\begin{figure}
    \centering
    \includegraphics[width=0.47\textwidth]{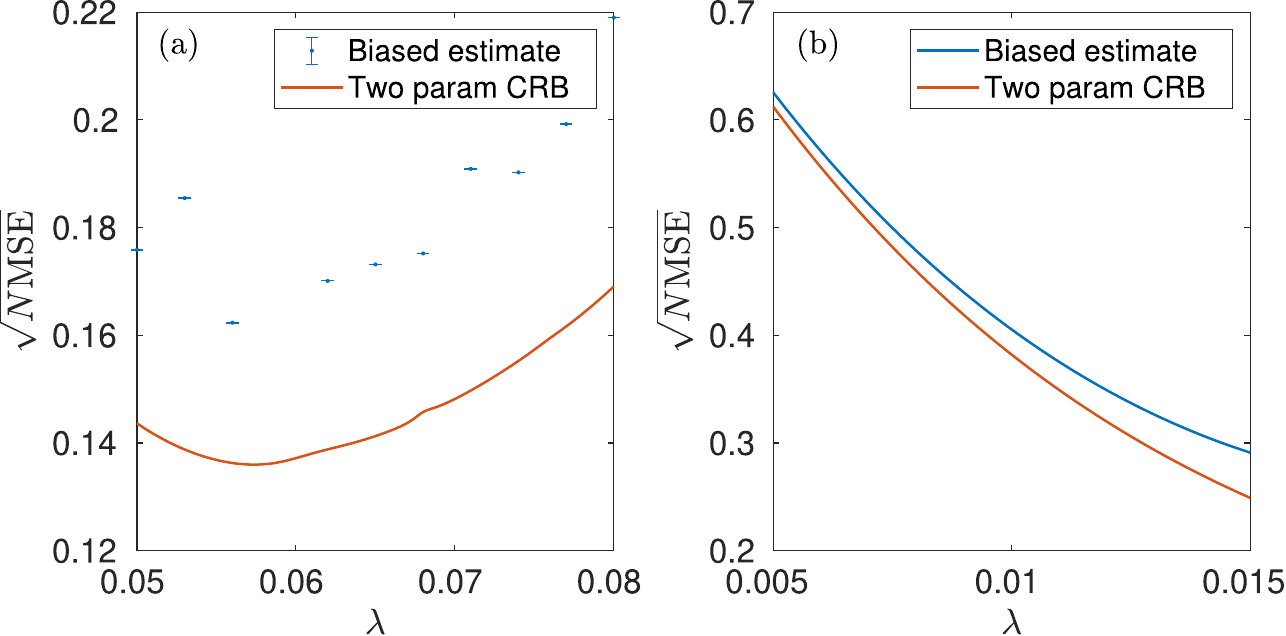}
    \caption{In (a), the normalised MSE for $\lambda'=\lambda+0.01$ for a TNT state, and in (b) the normalised MSE for a TAT state for $\lambda'=\lambda+0.005$, each for a sample size of $m=250$ and $\varphi=0.02$. The normalised two-parameter Cramér-Rao bound is plotted alongside the MSE. In both cases, we see that the two parameter method is superior to the MSE under a biased single parameter-estimator.}
    \label{fig:finalfig}
\end{figure}

We demonstrate that this two parameter estimation strategy can result in lower error for a sample size of $m=250$ in both the spin-squeezed and non-Gaussian regimes in Figure \ref{fig:finalfig}. We plot $\sqrt{N \text{MSE}}$ for TAT with from $\lambda=0.005$ to $\lambda=0.015$ for $\lambda'= \lambda+0.005$, and TNT for $\lambda=0.05$ to $\lambda=0.08$ and $\lambda'=\lambda+0.01$, alongside the two-parameter CRB. The TAT state is prepared according to Equation \ref{eqn:TATspinsqueezeunitary}, and the TNT state is prepared according to Equation \ref{eqn:TNTnonGaussianprotocol}. In Figure \ref{fig:finalfig} (a) and (b), we see precision for TNT and TAT under systematic state preparation error, and under two parameter maximum likelihood estimation, respectively. In both instances, we see that the two parameter estimate leads to increased precision. This indicates that using two parameter MLE where a state preparation parameter is not perfectly known is a viable method for precise quantum-enhanced atom interferometry, as well as a means of eliminating measurement bias.

\section{Conclusion}
We have shown that using entangled atom states in atom interferometry may undermine its most attractive feature - the ability to make calibration-free measurements. We derive analytic expressions for both the bias and mean-squared error under state preparation in the spin-squeezed regime, and a metric for local susceptibility to state preparation error under a maximum likelihood estimator that can be used in the non-Gaussian regime. This allows the exploration of the susceptibility of quantum states produced by important Hamilitonians such as OAT, TAT, and TNT, to state preparation error and to devise mathematical conditions on quantum states that are robust to state preparation error. We demonstrate how precision can counterintuitively be increased under a biased estimator in the absence of additional interferometer operations dependent on the state preparation parameter in the case of TAT. Finally, we present a means of retaining the accuracy of atom interferometry under state preparation error using two-parameter maximum likelihood estimation, which may come at the cost of precision for small sample sizes. 

\begin{acknowledgments}
We acknowledge the Ngunnawal people as the traditional custodians of the land upon which this research was conducted, and recognise that sovereignty was never ceded. We would like to acknowledge fruitful discussions had with Karandeep Gill, Zain Mehdi and John Close. SAH acknowledges support through an Australian Research Council Future Fellowship Grant No.~FT210100809.
\end{acknowledgments}

\bibliography{references}

\section{Appendixes}
In part A, we derive the expression for the quantum noise in a parameter of interest in a spin-squeezed state under state preparation error. In part B, we discuss how Kitagawa and Ueda's analytic expressions for OAT moments are used in Section 3 C. In part C, we derive an analytic expression for the bias of a maximum likelihood estimate under model misspecification. In part D, we briefly consider random, as opposed to systematic state preparation error, for TAT and OAT spin-squeezed states.

\label{app:bigeqder}
\subsection{Deriving Equation \ref{eqn:chap5var}}
We wish to calculate 
\begin{equation}
    \label{eqn:startappvar}
    \sigma_Q^2 = \dfrac{\text{Var}(\hat{J}_z)}{m| \partial_{\varphi} \langle \hat{J_z} \rangle|^2},
\end{equation}
in terms of the $\hat{J}_x$ rotation $\theta$, and the encoded parameter $\varphi$. We take Equation \ref{eqn:Jz expanded}, and compute the variance on the Heisenberg evolved operators as
\begin{widetext}
\begin{multline}
  \text{Var}(\hat{J}_z) = \cos^2(\theta)\cos^2(\varphi)\text{Var}(\hat{J}_{z_{0}})+\sin^2(\varphi)\text{Var}(\hat{J}_{x_0})
  +\cos^2(\varphi)\sin^2(\theta)\text{Var}(\hat{J}_{y_0})-\frac{1}{2}\sin(\theta)\sin(2\varphi)\overline{\text{Covar}}( \hat{J}_{x_0},\hat{J}_{y_0})\\
  -\frac{1}{2}\sin(2\varphi)\cos(\theta)\overline{\text{Covar}}(\hat{J}_{z_0},\hat{J}_{x_0})
  +\frac{1}{2}\sin(2\theta)\cos^2(\varphi)\overline{\text{Covar}}(\hat{J}_{z_0},\hat{J}_{y_0}),
\end{multline}
\end{widetext}
where $\overline{\text{Covar}}(a,b)=\text{Covar}(a,b)+\text{Covar}(b,a)$. Given that the initial spin-squeezed state is roughly symmetric about the $J_x$-axis, the covariance terms in $x,z$ and $x,y$ will be approximately zero. This, in combination with taking the expected value of Equation \ref{eqn:Jz expanded} allows us to write the standard error in $\varphi$ based on Equation \ref{eqn:startappvar} as
\begin{widetext}
\begin{multline}
    \sigma_Q^2 = \dfrac{1}{m|\cos(\varphi)\langle \hat{J}_{x_0}\rangle + \sin(\varphi)(\langle \hat{J}_{z_0}\rangle\cos(\theta) + \langle \hat{J}_{y_0}\rangle\sin(\theta))|^2} 
    \Bigg( \cos^2(\theta)\cos^2(\varphi)\text{Var}(\hat{J}_{z_0}) + \sin^2(\varphi)\text{Var}(\hat{J}_{x_0}) \\
    + \cos^2(\varphi)\left(\sin^2(\theta)\text{Var}(\hat{J}_{y_0}) + \frac{1}{2}\sin(2\theta)\overline{\text{Covar}}(\hat{J}_{z_0},\hat{J}_{y_0})\right) \Bigg).
\end{multline}
\end{widetext}

 By picking our initial state to be a CSS to be a $\hat{J}_x$ eigenstate as before, we can set $\langle \hat{J}_{y_0} \rangle =0$, $\langle \hat{J}_{z_0} \rangle =0$. Linearising around $\varphi=0$, we obtain
 
\begin{equation}
\begin{split}
    \label{equ:varnobias}
    \sigma_Q^2 &=  \dfrac{1}{m|\langle \hat{J}_{x_0}\rangle|^2} \left( \cos^2(\theta)\text{Var}(\hat{J}_{z_0})
    +\sin^2(\theta)\text{Var}(\hat{J}_{y_0}) \right.\\
    &\left. +\frac{1}{2}\sin(2\theta)\overline{\text{Covar}}(\hat{J}_{z_0},\hat{J}_{y_0}) \right).
\end{split}
\end{equation}
as desired.
\subsection{Analytic expressions for OAT moments}
\label{app:oatmom}
We use analytic expressions to evaluate the spin-moments of OAT developed by Kitagawa and Ueda \cite{Kitagawa1993}. Using $\bar{S}_i$ instead of $\hat{J}_i$ to denote a collective spin operator in a direction $i$, for $S=N/2,$ they found that
\begin{align*}
\left\langle\bar{S}_x\right\rangle &= S \cos ^{2 S-1} \frac{\mu}{2}, \quad \left\langle\bar{S}_y\right\rangle = 0, \quad \left\langle\bar{S}_z\right\rangle = 0, \\
\left\langle\Delta \bar{S}_x^2\right\rangle &= \frac{S}{2} \left[2 S \left(1-\cos ^{2(2 S-1)} \frac{\mu}{2}\right)-\left(S-\frac{1}{2}\right) A\right], \\
\begin{split}
\left\langle\Delta \bar{S}_{y,z}^2\right\rangle &= \frac{S}{2} \Bigg\{1 + \frac{1}{2} \left(S-\frac{1}{2}\right) \\
&\quad \times \left[A \pm \sqrt{A^2+B^2} \cos (2 \nu + 2 \delta)\right]\Bigg\}.
\end{split}
\end{align*}
where $\mu=2\lambda$, $A=1-\cos ^{2 S-2} \mu, B=4 \sin \frac{\mu}{2} \cos ^{2 S-2} \frac{\mu}{2}$, and $\delta=\frac{1}{2} \arctan \frac{B}{A}$. $\nu$ is the actual rotation angle about the $x$ axis, $\delta$ is the assumed angle that is applied to the OAT state in the simulation. The assumed angle $\nu$ applied in the simulation is $\pi-\delta'$, where $\delta$ is based on the assumed state preparation parameter $\lambda$.

\subsection{Derivation of analytic expressions for MLE bias}
\label{app:MLE_S}
Consider a likelihood function 
\begin{equation}
    \mathcal{L}_{\lambda_p}(\mathbf{X(\mathbf{\Lambda})} |\varphi) = \sum_{k} \log(P(X_k | \varphi, \lambda_p)), 
\end{equation}
where $\mathbf{X}$ represents a set of measurements dependent on the actual state of state parameters $\mathbf{\Lambda}$, $P$ represents a probability distribution parametrised by the estimator state preparation parameter $\lambda_p$, and $\varphi$ is the parameter of interest. We can expand $\mathcal{L}$ to second order around its maximum value $\varphi_0$ as
\begin{align*}
    \mathcal{L} &= \mathcal{L}(\lambda_p, \varphi_0) 
    + \dfrac{\partial \mathcal{L}}{\partial \lambda_p} (\lambda_p - \lambda_0) 
    + \dfrac{\partial \mathcal{L}}{\partial \varphi} (\varphi - \varphi_0) \\
    &\quad+ \dfrac{1}{2} \left( \dfrac{\partial^2 \mathcal{L}}{\partial \lambda_p^2} (\lambda_p - \lambda_0)^2 
    + \dfrac{\partial^2 \mathcal{L}}{\partial \varphi^2} (\varphi - \varphi_0)^2 \right. \\
    &\qquad\left. + 2 \dfrac{\partial^2 \mathcal{L}}{\partial \lambda_p \partial \varphi} (\lambda_p - \lambda_0)(\varphi - \varphi_0) \right).
\end{align*}

The derivative around the maximum with respect to $\varphi$ will be $0$, so we can write

\begin{equation}
\begin{split}
    \dfrac{\partial \mathcal{L}}{ \partial \varphi} = \dfrac{\partial^2 \mathcal{L}}{\partial \varphi^2}(\varphi - \varphi_0) + \dfrac{\partial^2\mathcal{L}}{\partial\lambda_p \partial\varphi}(\lambda_p - \lambda_0) = 0 
\end{split}
\end{equation}

from which we can calculate the bias as 

\begin{equation}
    \varphi - \varphi_0 = -\dfrac{\frac{\partial^2\mathcal{L}}{\partial\lambda_p \partial\varphi}}{\frac{\partial^2 \mathcal{L}}{\partial \varphi^2}}(\lambda_p - \lambda_0).
\end{equation}
The numerator and denominator turn out to be the off-diagonal and diagonal terms of the Fisher covariance matrix respectively, thus we can write the above equation as 
\begin{equation}
    \varphi - \varphi_0 = -\dfrac{F_{\lambda\varphi}}{F_{\varphi\varphi}}(\lambda_p - \lambda_0),
\end{equation}
so we can interpret the ratio of the off-diagonal and $\varphi$ elements of the Fisher covariance as a measure of the susceptibility of a quantum state in a particular basis to state preparation error.  

Given that $\lambda_p=\lambda'$, the assumed state preparation degree of freedom, and $\lambda_0=\lambda$, the actual state preparation degree of freedom, we can write that 
\begin{equation}
    \varphi^* - \varphi = \dfrac{F_{\lambda\varphi}}{F_{\varphi\varphi}}(\lambda'-\lambda).
\end{equation}

\subsection{Random state preparation error in spin-squeezed states}
Let us consider random state preparation error with multiple state preparation parameters $\mathbf{\Lambda}$, but only one degree of freedom $\lambda$ that fluctuates randomly with a known standard deviation $\Delta\lambda$. This $\Delta\lambda$ is defined separately from the $\Delta\lambda$ in the main text. The randomness in $\lambda$ results in the transition from a pure to a mixed state. After parameter encoding, the state before readout can be described as
\begin{equation}
\begin{aligned}
    \hat{\rho}_{\mathbf{\Lambda}, \varphi} &= \hat{U}_\varphi \int d\lambda \, \dfrac{1}{\sqrt{2\pi\Delta\lambda}} \\
    &\quad \times \exp\left( -\frac{(\lambda - \lambda_{0})^2}{2\Delta \lambda }\right) 
    \ket{\psi_{\mathbf{\Lambda}}}\bra{\psi_{\mathbf{\Lambda}}} \hat{U}^{\dagger}_\varphi,
\end{aligned}
\end{equation}

where $\ket{\psi_{\mathbf{\Lambda}}}\bra{\psi_{\mathbf{\Lambda}}}$ represents the density matrix of a pure quantum state entangled according to $\mathbf{\Lambda}$. Assuming that the experimenter has complete knowledge of the mixed state prepared, there will be no systematic error in the parameter estimate, i.e. $B=0$. 

We assume that our state preparation parameter vector is $\mathbf{\Lambda} = \{\lambda, \theta\}$,
where $\lambda$ is again the only degree of freedom.
 For a spin-squeezed state, the phase can be estimated based off $\langle \hat{J}_z\rangle$ as in Figure \ref{fig:systeestexplain} under an estimator accounting for the state preparation parameter fluctuations, and the variance can be calculated as
\begin{equation}
   \Delta \varphi= \dfrac{\Delta J_z}{\sqrt{m}|\partial_{\varphi}\mathbf{Tr}]\hat{\rho}_{\lambda}\hat{J_z}]|}.
   \label{eqn:stdphimixed}
\end{equation}

\begin{figure}
    \centering
    \includegraphics[width=0.95\columnwidth]{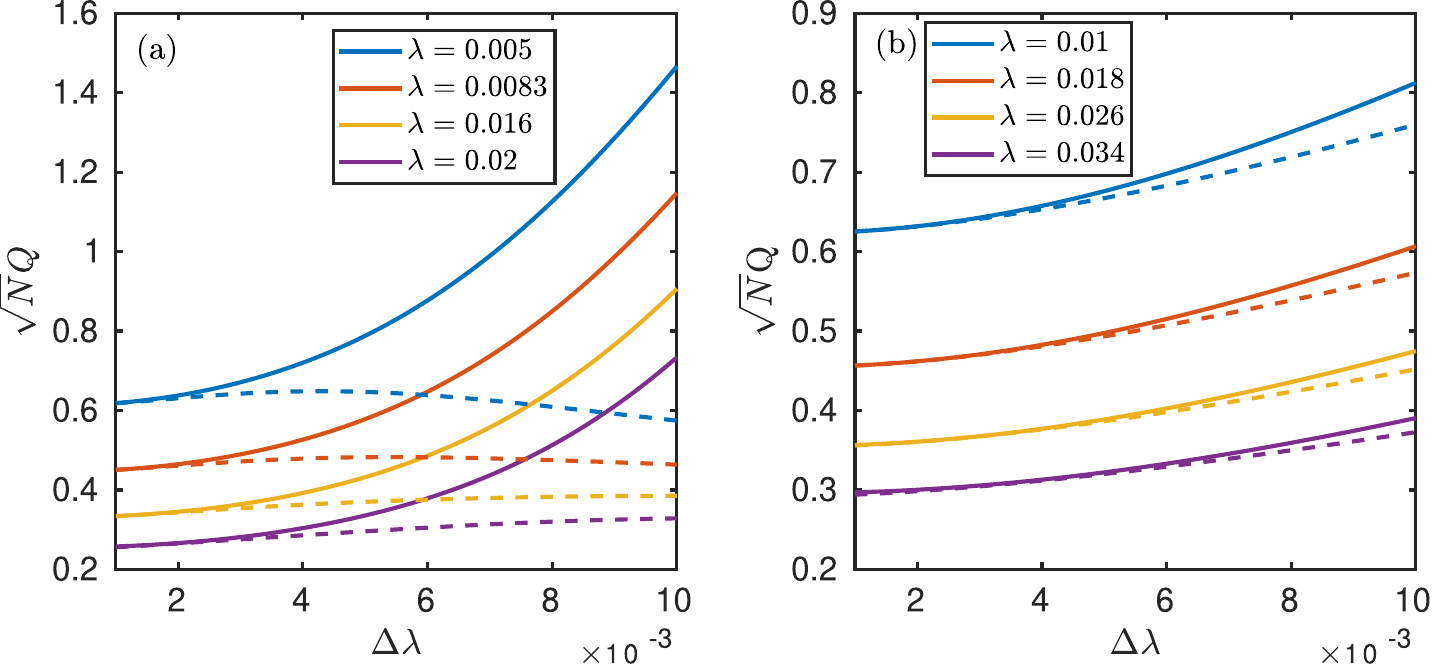}
    \caption{In (a) and (b), the scaled quantum noise for TAT and OAT states respectively for varying magnitudes of state preparation parameter $\Delta \lambda$ noise. The solid lines indicate the normalised $\varphi$ noise for a method of moments estimator, while the dotted lines in the same colour indicate the phase sensitivity of the state according the CRB, i.e. the sensitivity attainable with a maximum likelihood estimate. In both (a) and (b), we see that MSE increases monotonically for increasing state preparation parameter fluctuations. This is due to the increase of variance in the pseudo-spin operators as a result of a mixture of a wide variety of states. In (a) for TAT, we see that a maximum likelihood estimate diverges from the MOM estimate due to our mixed state no longer being Gaussian, and we see higher $\Delta \lambda$ leading to lower noise in some cases as higher CFI states are mixed in. In (b) for OAT, the maximum likelihood estimate improves on the MOM estimate, but $Q$ is monotonically increasing, as all but one state in the OAT fixture are rotated by a suboptimal angle, meaning comparatively lower CFI states.}
    \label{fig:mixed_state_fig}
\end{figure}

We consider the sensitivity ratio for varying levels of mixed state preparation error for TAT and OAT. In the case of OAT, we apply the optimal $\hat{J}_x$ rotation for the mean state preparation value in the distribution, $\lambda_0$. We find that increased uncertainty in the $\lambda$ value leads to greater state preparation error for both OAT and TAT under a method of moments estimator. In Figure \ref{fig:mixed_state_fig} (a) for TAT and (b) for OAT, as $\Delta \lambda$ increases, the variance in $\hat{J}_x$ increases.  This is because the state's total variance takes the possible range of $J_z$ and $J_x$ distributions given by the randomly distributed $\lambda$. Given that the state is squeezed along the $J_z$ axis, the variance of the mixed state for all fixed values of $\lambda_0$ will be greater in $\hat{J}_x$ than in $\hat{J}_z$, and monotonically increasing in both, leading to an increase in error according to Equation \ref{eqn:chap5var}.

In the cases of both OAT and TAT, we see that a maximum likelihood estimation strategy improves on a method of moments strategy for $\Delta \lambda \neq 0$. This is because more highly mixed states tend away from being Gaussian, where the quantum state can be characterised by a single moment, and thus the CRB can be saturated by a MOM estimator. 
When using a maximum likelihood estimation strategy, we see a distinction between OAT and TAT. In the case of TAT, increased uncertainly in $\lambda$ does not necessarily decrease precision. This is because as $\lambda$ increases, less spin-squeezed state are mixed in with more highly spin-squeezed states. The $\hat{J}_z$ distribution becomes weighted with the more highly distinguishable distributions of spin-squeezed states, resulting in a distribution that is overall more distinguishable, thus having higher Fisher information than mixed states with lower $\Delta \lambda$ values. In the case of OAT on the other hand, we find increased state preparation uncertainty leads to uniformly increasing $Q$. This is because the phase space rotation after squeezing varies for the state preparation parameter, unlike for TAT. The varied state preparation parameter results in a mixture of states that are not all squeezed along the same axis, as the optimal $\hat{J}_x$ rotation can not be applied for all states - far fewer highly distinguishable distributions become part of the new probability distribution.

\end{document}